\documentclass[pra,aps,showpacs,floats,twocolumn]{revtex4-1}
\usepackage{graphicx}

\begin{document}

\title{Mott transitions in a three-component Falicov-Kimball model: \\A slave boson mean-field study}

\author{Duc-Anh Le$^{1}$ and Minh-Tien Tran$^{2,3}$}
\affiliation{$^1$Faculty of Physics, Hanoi National University of Education,136 Xuan Thuy, Cau Giay,  Hanoi, Vietnam}
\affiliation{$^2$Institute of Research and Development, Duy Tan University, K7/25 Quang Trung, Danang, Vietnam}
\affiliation{$^3$Institute of Physics, Vietnam Academy of Science and Technology, 10 Dao Tan, Hanoi, Vietnam}


\begin{abstract}
Metal-insulator transitions in a three-component
Falicov-Kimball model are investigated within the Kotliar-Ruckenstein slave boson mean-field approach.
The model describes a mixture of two interacting fermion atom species loaded into an optical lattice at ultralow temperature. One species is two-component atoms, which can hop in the optical lattice, and the other is single-component atoms, which are localized. Different correlation-driven metal-insulator transitions are observed depending on the atom filling conditions and local interactions. These metal-insulator transitions are classified by the band renormalization factors and the double occupancies of the atom species. The filling conditions and the critical value of the local interactions for these metal-insulator transitions are also analytically established. The obtained results not only are in good agreement with the dynamical mean-field theory for the three-component Falicov-Kimball model but also clarify the nature and properties of the metal-insulator transitions in a simple physics picture.
\end{abstract}
\maketitle

\section{\label{sec:1} Introduction}
Recently, Mott transitions in multicomponent systems have attracted a lot of attention \cite{Hofstetter,Gorshkov,Hermele,Gorelik,Inaba}. They give rise to intriguing and rich physics that does not have obvious counterparts in physics of condensed matters. The multicomponent systems are realized by cooling neutral atoms, for example, multicomponent lithium at ultra-low temperature \cite{Ottenstein,Huckans}. When these multicomponent systems are loaded into optical lattices, they simulate various theoretical models of electron correlations with extensions of degrees of freedom, such as the multicomponent Hubbard and Heisenberg models \cite{Bloch}. The simulations provide a connection between the theoretical studies and experimental observations. Indeed, experiments already found the theoretically predicted Mott insulating phase in two-component fermion systems which simulate the single-band Hubbard model \cite{Jordens,Schneider}. In three-component fermion systems
the Mott insulating phase still exists at commensurate fillings, and in addition also at incommensurate fillings, depending on the symmetry of the local interactions between the particle components \cite{Gorelik,Inaba}.

In a parallel development, experiments have also realized multicomponent atom mixtures that pave the way to the investigation of the Mott insulating states of different particle species \cite{Tag,Spiegelhalder,Taie,Hara}. This investigation has also attracted a lot of theoretical studies. Various phase transitions have been theoretically discussed for different kind of mixtures, including boson-boson \cite{Hubener,Sansone}, boson-fermion \cite{Tit,Lewenstein}, and fermion-fermion mixtures \cite{Iskin,Bo2013}. In fermion-fermion mixtures, different metal-insulator transitions (MITs), which include collective, species selective, and inverse Mott transitions, were theoretically found \cite{Bo2013}. The finding is based on
the dynamical mean-field theory (DMFT)~\cite{Metzner,GKKR} for a three-component Falicov-Kimball model (FKM), that describes a fermion-fermion mixture of two-component light atoms and single-component heavy atoms~\cite{Bo2013}.
This model is an extension with odd number of components of the spinless FKM, which exhibits a rich phase diagram and has also attracted a lot of studies \cite{FKM,FreericksZlatic}. It can also be considered as a mass balanced case of the three-component Hubbard model \cite{Hofstetter,Gorshkov,Hermele,Gorelik,Inaba}. One may expect
rich phase transitions in the three-component FKM, that reflect the odd and non-equivalent component multiplicity and do not have obvious counterparts in two-component or three equivalent component systems. Indeed,
at commensurate fillings both the inter- and intra-species interactions collectively drive the mixture from the metallic to the insulating state, and at the incommensurate half filling, only two-component light atoms are involved in the Mott transition. The species-selective MIT is reminiscent of the orbital-selective one, where the narrow orbital band becomes insulating, while the wide band is still metallic \cite{Anisimov,Koga,Liebsch2004,Medici}. When the intra-species interaction is weaker than the inter-species one, the correlations between atom components can also drive the system from the insulating to the metallic state \cite{Bo2013}.

The aim of the present paper is two-fold. First, in the present paper, the Mott transitions in the three-component Falicov-Kimball model are studied by the  Kotliar-Ruckenstein slave boson mean-field approach \cite{Kotliar,Zwickgnal,Sigrist, anh2012,anh2014}.
The previous work  has employed the DMFT plus exact diagonalization (ED)~\cite{Bo2013}. Although the ED is an exact impurity solver, it is limited to a finite number of the energy levels of the dynamical mean field and always admits finite-size effects.
Within the DMFT the double occupancy always remains finite even in the Mott insulator state and it cannot be used as an order parameter of the MIT \cite{GKKR,Tien}.
In addition, within the DMFT, the dynamics of the localized heavy fermions is complicated to calculate, because it is completely excluded from the effective mean-field single impurity dynamics \cite{FreericksZlatic,Brandt1}.
In multi-orbital systems, different impurity solvers of the DMFT often capture different aspects of the Mott transitions, and can partially lead to different conclusions \cite{Anisimov,Koga,Liebsch2004,Medici}.
It is thus desirable to investigate the properties of these Mott transitions within a more analytical trackable theory.
Therefore, we apply the Kotliar-Ruckenstein slave boson approach~\cite{Kotliar,Zwickgnal,Sigrist,anh2012,anh2014} on the mean-field level to the three-component FKM.
Our reasonable results not only agree well with the DMFT~\cite{Bo2013} but also clarify the nature of the Mott transitions in a simple physics picture through the band renormalization factors and the double occupancies. In addition, the slave boson mean field approximation also allows us to determine not only the dynamics of itinerant fermions, but also the dynamics of localized fermions, that is complicated to calculate within the DMFT. The second aim of the present paper is to show that the Kotliar-Ruckenstein slave boson approach at the mean field level already fully captures  the description of the MIT in the multicomponent fermion mixtures, and it can even be analyzed in an analytical manner. Since the slave boson mean field approximation is quite simple, and its description of the MIT is physically intuitive, this approach serves an alternative method of investigating multicomponent correlated systems.

The structure of the paper is as follows. In Sec. \ref{sec2} we present the model as well as introduce the auxiliary boson representation in the mean-field approximation. In Sec. \ref{sec3} we present and discuss the numerical results of the ground state in paramagnetic states at both commensurate and incommensurate fillings. Finally, the conclusions are presented in Sec. \ref{sec4}.

\section{\label{sec2} Three-component Falicov-Kimball model and the Kotliar-Ruckenstein slave boson approach}
We consider the following Hamiltonian of a three-component FKM that describes a mixture of
heavy and light fermion atoms loaded in an optical lattice~\cite{Bo2013}
\begin{eqnarray}
H &=& -J\sum_{<i,j>,\sigma} c^{\dagger}_{i\sigma} c_{j\sigma}-\mu_c \sum_{i,\sigma} c^{\dagger}_{i\sigma} c_{i\sigma} - \mu_f \sum_{i} f^{\dagger}_{i} f_{i}\nonumber \\
&&+ U_{cc}\sum_{i}c^{\dagger}_{i\uparrow} c_{i\uparrow} c^{\dagger}_{i\downarrow}c_{i\downarrow} + U_{cf} \sum_{i,\sigma} f^{\dagger}_{i} f_{i}  c^{\dagger}_{i\sigma} c_{i\sigma},  \label{ham}
\end{eqnarray}
where $c^{\dagger}_{i\sigma}$ ($c_{i\sigma}$) is the creation
(annihilation) operator for two-component fermion atoms at lattice site $i$ ($\sigma\equiv\uparrow,\downarrow$), and
$f^{\dagger}_{i}$ ($f_{i}$) is the creation (annihilation)
operator for single-component fermion atoms at
lattice site $i$.
$J$ is the nearest neighbor hopping amplitude of the two-component
fermion atoms. $\mu_c$ and $\mu_f$ are the chemical potentials of the two- and single- component atoms, respectively.
They control the
particle species fillings $n_{c\sigma}=\sum_{i} \langle c^{\dagger}_{i\sigma} c_{i\sigma}
\rangle /N$ and $n_{f}=\sum_{i} \langle f^{\dagger}_{i} f_{i}
\rangle /N$, where $N$ is the number of lattice sites.
 $U_{cc}$ is the intra-species local repulsive interaction between
the two-component atoms, and
 $U_{cf}$ is the inter-species one.
The Hamiltonian in Eq.~(\ref{ham}) describes a fermion-fermion mixture, where only the two-component atoms are able to hop in the lattice, and
the single-component atoms are extremely heavy and are always localized. Actually, the hopping amplitude of atoms in optical lattices is tuned by the optical potential and the recoil energy of each atom species of the mixture \cite{Zwerger}. With sufficiently deep potential, the energy band of atoms in optical lattices  can become flat and the hopping amplitude vanishes. The mixture can be considered as an extreme mass imbalance case of the three-component fermion mixtures \cite{Gorelik,Inaba}. The three-component FKM can also be considered as  a multicomponent extension of the spinless FKM~\cite{FKM} or an asymmetric simplification of the three-component Hubbard model \cite{Gorelik}. It has two well-known limiting cases. When $U_{cc}=0$, the three-component FKM is reduced to the spinless FKM, which exhibits a Mott-like MIT at half filling \cite{Dongen92,Dongen}. When $U_{cf}=0$, it is equivalent to the single-band Hubbard model, which also exhibits the Mott MIT at half filling \cite{Hubbard}. However, when both $U_{cc}$ and $U_{cf}$ are finite, different MIT may occur in the three-component FKM, that reflect the component multiplicity and do not have obvious counterparts of the limiting cases \cite{Bo2013}.

\begin{table}
\caption{Local states ($|\Gamma\rangle$), their energy levels ($E_\Gamma$) and corresponding slave boson ($\phi_\Gamma$) representation.}
\begin{tabular}{|c|c|c|c|r|}
\hline
$\Gamma$ & $\vert \Gamma\rangle$  &
$E_{\Gamma}$ &$\phi^{\dagger}_{\Gamma}$& {Slave boson representation}\\
\hline
1 & $\vert 0\rangle$ &  $0$ &  $e^{\dagger}$ &$e^{\dagger}\vert 0\rangle$\\
2   & $c^{\dagger}_{\uparrow}\vert 0\rangle$ & $0$ &  $p^{\dagger}_{\uparrow}$ &$\hat{c}^{\dagger}_{\uparrow}p^{\dagger}_{\uparrow}\vert 0\rangle$\\
3   & $c^{\dagger}_{\downarrow}\vert 0\rangle$ & $0$ &  $p^{\dagger}_{\downarrow}$ &$\hat{c}^{\dagger}_{\downarrow}p^{\dagger}_{\downarrow}\vert 0\rangle$ \\
4   & $f^{\dagger}\vert 0\rangle$ & $0$ &  $p^{\dagger}_{f}$ &$\hat{f}^{\dagger}p^{\dagger}_{f}\vert 0\rangle$\\
5  & $c^{\dagger}_{\uparrow}c^{\dagger}_{\downarrow}\vert 0\rangle$ & $U_{cc}$ &  $d^{\dagger}_{\uparrow\downarrow}$ &$\hat{c}^{\dagger}_{\uparrow}\hat{c}^{\dagger}_{\downarrow}d^{\dagger}_{\uparrow\downarrow}\vert 0\rangle$\\
6 & $c^{\dagger}_{\uparrow}f^{\dagger}\vert 0\rangle$ &$U_{cf}$ &  $d^{\dagger}_{\uparrow f}$ &$\hat{c}^{\dagger}_{\uparrow}\hat{f}^{\dagger}d^{\dagger}_{\uparrow f}\vert 0\rangle$\\
7 & $c^{\dagger}_{\downarrow}f^{\dagger}\vert 0\rangle$  & $U_{cf}$ &  $d^{\dagger}_{\downarrow f}$ &$\hat{c}^{\dagger}_{\downarrow}\hat{f}^{\dagger}d^{\dagger}_{\downarrow f}\vert 0\rangle$\\
8 & $c^{\dagger}_{\uparrow}c^{\dagger}_{\downarrow}f^{\dagger}\vert 0\rangle$& $U_{cc}+2U_{cf}$ &  $t^{\dagger}_{}$&$\hat{c}^{\dagger}_{\uparrow}\hat{c}^{\dagger}_{\downarrow}\hat{f}^{\dagger}t^{\dagger}\vert 0\rangle$ \\
\hline
\end{tabular}
\end{table}

We generalize the Kotliar-Ruckenstein slave boson representation~\cite{Kotliar} for the three-component FKM described in Eq. (\ref{ham}).
Within the Kotliar-Ruckenstein slave boson representation, every local state at each lattice site is represented by an auxiliary (slave) boson \cite{Kotliar,Zwickgnal,Sigrist,anh2012,anh2014}.
Since each lattice site of the three-component FKM can be empty, singly, doubly, or triply occupied, we use eight slave bosons
$e^{\dagger}(i)$, $p^{\dagger}_{\sigma}(i)$, $p^{\dagger}_{f}(i)$, $d^{\dagger}_{\uparrow\downarrow}(i)$, $d^{\dagger}_{\uparrow f}(i)$, $d^{\dagger}_{\downarrow f}(i)$, $t^{\dagger}(i)$ to represent these eight different local states.
The notations of the slave bosons $e$, $p$, $d$, $t$ denote the empty, singly, doubly and triply occupied states, respectively.
For convenient, the eight slave bosons are labeled by numbers as shown in table 1, e.g., $\phi^{\dagger}_1(i)  = e^{\dagger}(i)$, $\phi^{\dagger}_2(i)  = p^{\dagger}_{\uparrow}(i)$, etc. In the table 1 for the sake of clarity, we skip the site index.
For the three fermion components we use three auxiliary fermions $\hat{c}_{i\sigma}$ and $\hat{f}_{i}$.
The original fermion operators  are expressed
in term of auxiliary slave bosons and fermions as follows~\cite{Kotliar,Zwickgnal,Sigrist,anh2012,anh2014}
\begin{eqnarray}
c_{\sigma} &=&
\hat{R}_{\sigma}^{\dagger}[\Phi]\hat{c}_{\sigma}, \\
f &=&
\hat{R}_{f}^{\dagger}[\Phi]\hat{f} ,
\end{eqnarray}
where the operator $\hat{R}_{\alpha}^{\dagger}[\Phi]$ is defined as
\begin{eqnarray}
\hat{R}_{{\alpha}}^{\dagger}[\Phi]\equiv
\frac{\hat{\gamma}_{\alpha}[\Phi]}{\sqrt{\hat{n}_{\alpha}[\Phi]
\left(1-\hat{n}_{\alpha}[\Phi]\right)}},
\label{eq:QPmatrix}
\end{eqnarray}
with
\begin{eqnarray}
\label{Defgamma}
\hat{\gamma}_{\alpha}[\Phi]&\equiv&\sum_{\Gamma\Gamma'}
 |\langle \Gamma|a^{\dagger}_{\alpha}|\Gamma'\rangle|^2
\phi^{\dagger}_{\Gamma}\phi_{\Gamma'},\\
\label{Defnjz}
\hat{n}_{\alpha}[\Phi]
&\equiv& \sum_{\Gamma}\phi^\dagger_{\Gamma}\phi_{\Gamma}
\langle \Gamma |a^\dagger_{\alpha} a_{\alpha}|\Gamma\rangle.
\end{eqnarray}
Here we used the notation $a^{\dagger}_{\alpha}$ to denote either $c^{\dagger}_{\sigma}$ or $f^{\dagger}$, $\alpha=\sigma,f$, and skipped the lattice site index.
The introduced slave bosons already  enlarge the local state space at each lattice site. Therefore, we should eliminate
the unphysical states by imposing the local constraint
\begin{equation}
\sum_{\Gamma}\phi^\dagger_{\Gamma}(i)
\phi_{\Gamma}(i)= 1 .
\label{constraints1}
\end{equation}
This constrain is the closure relation of the local states at each lattice site. The other constraints which must be also imposed are
the identities of the fermion numbers counted through the auxiliary fermions and original fermions
\begin{eqnarray}
\sum_{\Gamma}\phi^\dagger_{\Gamma}(i)\phi_{\Gamma}(i)
\langle \Gamma(i) |c^\dagger_{i\sigma} c_{i\sigma}|\Gamma(i)\rangle&=&\hat{c}^\dagger_{i\sigma} \hat{c}_{i\sigma},
\label{constraints2a}\\
\sum_{\Gamma}\phi^\dagger_{\Gamma}(i)\phi_{\Gamma}(i)
\langle \Gamma(i) |f^\dagger_{i} f_{i}|\Gamma(i)\rangle&=&\hat{f}^\dagger_{i} \hat{f}_{i}.
\label{constraints2}
\end{eqnarray}
The constrains in Eqs. (\ref{constraints1})-(\ref{constraints2}) must be taken into account by introducing the Lagrange multipliers.

Hamiltonian in Eq.~(\ref{ham}) without the chemical potential terms now can be rewritten in terms of auxiliary fermions
and bosons as~\cite{Zwickgnal,Sigrist,anh2012,anh2014}
\begin{eqnarray}
H &=& -J\sum_{\langle i,j\rangle,\,\sigma}
\left[
\hat{R}_{\sigma}[\Phi (i)]
\hat{R}_{\sigma}^\dagger[\Phi (j)]
\hat{c}^\dagger_{i \sigma} \hat{c}_{j \sigma}
+
{\rm H. c.}\right]\nonumber\\
&&  +
\sum_{i,\,\Gamma} E_{\Gamma}
\phi^\dagger_{\Gamma}(i)
\phi_{\Gamma}(i) .
\label{eq:auxiliaryHalmintonian}
\end{eqnarray}
Actually, the chemical potential terms can be absorbed into the constrain terms (\ref{constraints2a})-(\ref{constraints2}). Within the Kotliar-Ruckenstein slave boson representation, the local interaction terms become quadratic of the slave bosons. This feature is a benefit of the slave boson approach on the cost of a complication of the hopping term and additional constraints.
The single-component atoms are localized, their effects only enter through the constraint term in Eq. (\ref{constraints2}) like the chemical potential term. However, the correlation effects still persist with the single-component atoms through the slave bosons.
At the mean-field level, the slave bosons are replaced by c-numbers and in the homogeneous phases they can be assumed to be site-independent. This greatly simplifies calculations and allows us to perform a trackable analysis. We use the following notations: $e$, $p_{\sigma}$, $p_{f}$, $d_{\uparrow\downarrow} $, $d_{\uparrow f}$, $d_{\downarrow f}$, $t$ for the mean-field value of the slave bosons $e^{\dagger}(i)$, $p^{\dagger}_{\sigma}(i)$, $p^{\dagger}_{f}(i)$, $d^{\dagger}_{\uparrow\downarrow}(i) $, $d^{\dagger}_{\uparrow f}(i)$, $d^{\dagger}_{\downarrow f}(i)$, $t^{\dagger}(i)$, respectively.  At zero temperature we obtain the ground-state energy per site~\cite{Zwickgnal,Sigrist,anh2012,anh2014}
\begin{equation}
E = -\sum \limits_{\sigma}\frac{W}{2} \gamma_{\sigma}[\Phi]^2 + \sum_{\Gamma} E_{\Gamma}
\phi^2_{\Gamma} .
\label{EG}
\end{equation}
Here for simplicity, a constant bare density of states $\rho_0(\omega) = \frac{1}{W} \theta \left( {\frac{W}{2}-\vert \omega \vert}\right )$ with the bandwidth $W$ has been used.
In order to find the  mean-field ground state, Eq.~(\ref{EG}) is minimized with the constraints (\ref{constraints1})-(\ref{constraints2}), which are also in the mean field approximation. Solving the minimization equations, we obtain the mean-field values of the slave bosons as well as the particle fillings. After that
we calculate the band renormalization factor $Z_\alpha$, the intra-species double occupancy $D_{cc}$ and the inter-species double occupancy $D_{\sigma f}$ which are defined as
\begin{eqnarray}
Z_\alpha = R^2_{\alpha}[\Phi]=\frac{\gamma^2_{\alpha}}{n_{\alpha}\left( 1 - n_{\alpha}\right)},
\label{rf} \\
D_{cc}\equiv \langle n_{\uparrow}n_{\downarrow} \rangle = d^2_{\uparrow \downarrow} + t^2, \label{docc}\\
D_{\sigma f} \equiv  \langle n_{\sigma}n_{f}\rangle = d^2_{\sigma f} + t^2. \label{docf}
\end{eqnarray}
The interactions between particles renormalize the effective mass of the particles by the band renormalization factor. When the band renormalization factor vanishes, the effective mass becomes infinite and the particles become localized. This is the standard Brinkman-Rice scenario of the Mott insulating state \cite{Brinkman}. Although the single-component atoms are localized, the local interactions still renormalize their bare energy level by the band renormalization factor $Z_f$.  Within the DMFT the self energy of the localized particles is complicated to calculate, because their dynamics is completely excluded from the effective single impurity dynamics of the dynamical mean field \cite{FreericksZlatic,Brandt1}. By using the Kotliar-Ruckenstein slave boson approach, the band renormalization factors of both itinerant and localized atoms are determined on the same footing.
The band renormalization factors, the intra- and the inter-species occupancies exhibit distinct behaviors in the metallic and the insulating states. Therefore, we can use them to monitor the MIT. Actually, experiments of ultracold atoms loaded in optical lattices have also detected the Mott insulating state by counting the doubly occupied sites \cite{Jordens}. In this work, we restrict ourself to the paramagnetic phase, however, the calculations for magnetically ordered phases are straightforward. Note that, in the paramagnetic phase, one has $p_{\uparrow}=p_{\downarrow}\equiv p_c$,
$d_{\uparrow f}=d_{\downarrow f}\equiv d_{cf}$, $n_{c\uparrow}=n_{c\downarrow}\equiv n_c$, $D_{\uparrow f} = D_{\downarrow f} = D_{cf} $ and $Z_{\uparrow}=Z_{\downarrow}\equiv Z_c$.
\section{\label{sec3} Numerical results and discussions}
In this section we present the numerical results obtained by minimizing the ground-state energy (\ref{EG}) with the constraints (\ref{constraints1})-(\ref{constraints2}). We obtain $8+3$ nonlinear equations.
The main difficulty here is how to reach the saddle point
efficiently in the $8+3$ dimensional parameter space. The simple iterative procedure is very hard to be converged and it is
almost useless in practice. Instead, we use a modification of the Powell hybrid method.~\cite{Powell} This algorithm is a variation of Newton's method, which takes precautions to avoid large step sizes or increasing residuals. In the numerical calculations we take the bare bandwidth $W=1$ as the unit of energy.

\subsection{Mott transitions and their filling conditions}
\begin{figure}[t]
\begin{center}
\includegraphics[width=0.4\textwidth]{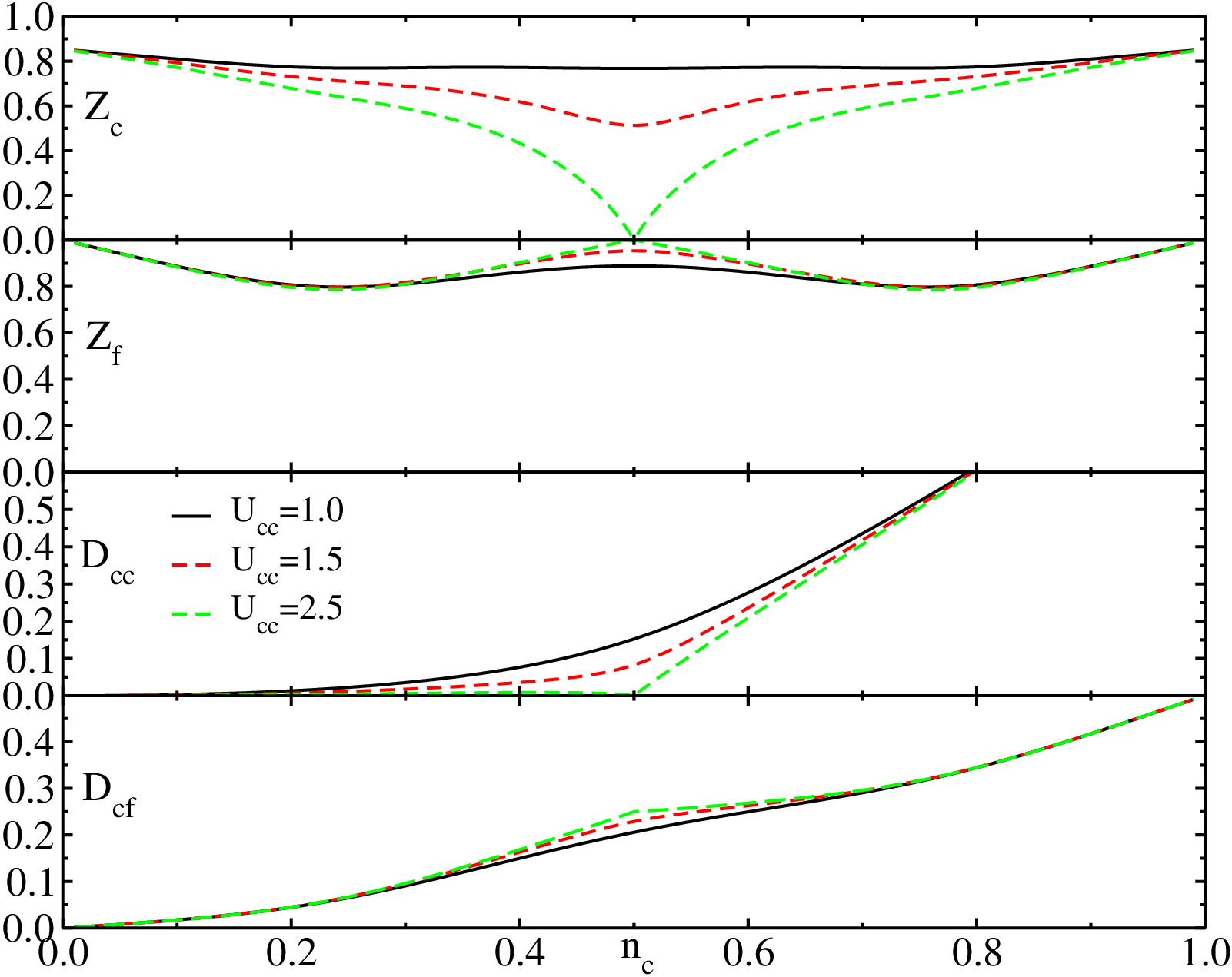}
\caption{(Color online) The band renormalization factors $Z_c$ and $Z_f$ and the intra- $D_{cc}$ and the inter-$D_{cf}$
species double occupancies as a function of the two-component atom filling $n_{c}$ in the weak inter-species interaction region ($U_{cf}=0.5$)
at the single-component atom filling $n_f=1/2$.}
\label{fig1}
\end{center}
\end{figure}

In this subsection we determine the particle filling conditions, where the Mott transition may occur. We find different kinds of the  Mott transition depending on the local interactions and the particle fillings \cite{Bo2013}.
First, we consider the region of weak inter-species interactions ($U_{cf} < W$).
In Fig. \ref{fig1} we plot the band renormalization factors of both atoms species, as well as the intra- and inter-species double occupancies as a function of two-component atom filling $n_{c}$ for a given weak inter-species local interaction at fixed single-component atom filling $n_f=1/2$. One can see that for weak intra-species local interactions $U_{cc}$, the band renormalization factor $Z_c$ and the double occupancy $D_{cc}$ of the two-component atoms are finite for all fillings $n_c$. However, at strong intra-species local interactions $U_{cc}$, they together vanish at filling $n_c=1/2$. This is a signal of the localization of the two-component atoms. However, the single-component atoms exhibit different behaviors.
Figure \ref{fig1} also shows that the band renormalization factor $Z_f$ of the single-component atoms
is always finite for all fillings $n_c$. In contrast to the two-component atoms,
the band renormalization factor $Z_f$ of the single-component atoms approaches to $1$ at filling $n_c=1/2$, when the intra-species local interaction $U_{cc}$ becomes strong.
The inter-species double occupancy $D_{cf}$ also equals to the noninteracting value $1/4$ at filling $n_c=1/2$.
These features indicate that at half filling the single-component atoms behave like free ones at strong intra-species local interactions.
The vanishing of the band renormalization factor $Z_c$ and the intra-species double occupancy $D_{cc}$ of the two-component atoms, while the single-component atoms become free at filling $n_c=1/2$, suggests that the MIT at filling $n_c=1/2$ deals only with the two-component atoms.
This MIT is referred to as a species-selective one \cite{Bo2013}.
In the species-selective MIT, only the two-component atoms are involved in the forming of the Mott insulating state.  This is reminiscent of the orbital-selective MIT in the multi-orbital systems, where only the narrow orbital band becomes insulating, while the broad bands still remain metallic \cite{Anisimov,Koga,Liebsch2004,Medici}. However, in the multi-orbital systems, all bands are always renormalized by interactions, and the orbital-selective MIT requires a finite Hund coupling. In this species-selective MIT, the single-component atoms are free of the interaction renormalization independently of the local interactions and filling $n_f$. Actually, in the insulating state, the particle number fluctuations of the two-component atoms are suppressed; the inter-species interaction acts like a shift of the chemical potential for the single-component atoms. As a result, the single-component atoms become effectively free in the insulating state irrespective of the local interactions and single-component atom filling $n_f$.
The freedom of the single-component atoms in the Mott insulator is an unique feature of the species-selective MIT. However, the effect of the inter-species local interaction still persists. It reflects on the dependence of the critical value of the intra-species local interaction, where the MIT occurs, on the inter-species local interaction.

\begin{figure}[t]
\begin{center}
\includegraphics[width=0.4\textwidth]{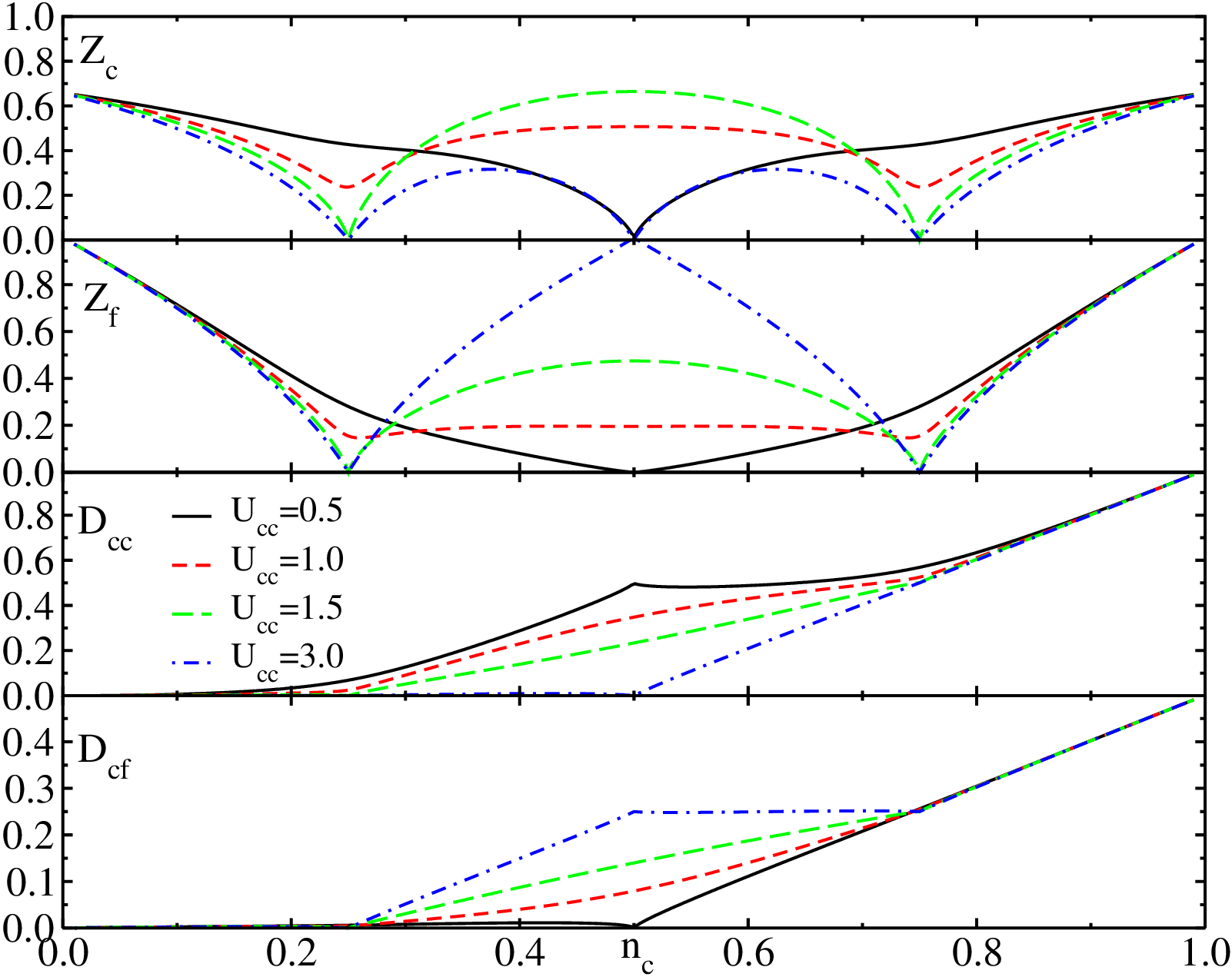}
\caption{(Color online) The band renormalization factors $Z_c$, $Z_f$ and the intra- $D_{cc}$ and the inter- $D_{cf}$
species double occupancies as a function of the two-component atom filling $n_{c}$ in the strong inter-species interaction region ($U_{cf}=1.5$)
at the single-component atom filling $n_f=1/2$.}
\label{fig2}
\end{center}
\end{figure}

Next, we consider the region of strong inter-species interactions ($U_{cf}>W$). In this region, we observe different MITs.
In Fig.~\ref{fig2} we plot the band renormalization factors of both atoms species, as well as the intra- and inter-species double occupancies as a function of two-component atom filling $n_{c}$ for a given strong inter-species local interaction at fixed single-component atom filling $n_f=1/2$. One can see the vanishing of the band renormalization factor $Z_c$ of the two-component atoms at different fillings $n_c$. First, we still observe the vanishing of $Z_c$ at filling $n_c=1/2$. However, at filling $n_c=1/2$ the band renormalization factor $Z_c$ vanishes at both weak and strong intra-species local interactions, for instance, at $U_{cc}=0.5$ and $3.0$ as shown in Fig. \ref{fig2}. In contrast to the species-selective MIT, for weak intra-species local interactions, the intra-species double occupancy $D_{cc}$ remains finite. The vanishing of the band renormalization factor $Z_c$ indicates the localization of the two-component atoms, but their weak intra-species local interaction does not prevent their double occupation. On the other hand,
the band renormalization factor $Z_f$ of the single-component atoms and the inter-species double occupancy $D_{cf}$ vanish at filling $n_c=1/2$ for weak intra-species local interactions. When the intra-species local interaction is larger than a certain value, both the band renormalization factors $Z_c$ and $Z_f$ stop vanishing. This indicates a transition from insulating to metallic states. This MIT is referred to as an inverse MIT, where particle correlations drive the mixture from insulator to metal \cite{Bo2013}. For other fillings $n_f$ we still observe the inverse MIT when $n_c+n_f=1$ is fulfilled. This MIT is similar to the Mott-like transition in the spinless Falicov-Kimball model \cite{FreericksZlatic}, where both band renormalization factors $Z_c$ and $Z_f$, and the double occupancy $D_{cf}$ vanish in the insulating state. 
Figure \ref{fig2} also shows the vanishing of the band renormalization factor $Z_c$ of the two-component atoms at filling $n_c=1/2$ for strong intra-species local interactions $U_{cc}$. In this case $D_{cc}=0$ and $Z_f=1$. These features constitute the species-selective MIT, which we have discussed previously.

In Fig.~\ref{fig2} we also observe the vanishing of the band renormalization factor $Z_c$ of the two-component atoms at other fillings, $n_c=1/4$ and $3/4$, when the local interactions are strong. At filling $n_c=1/4$, the band renormalization factors of both atom species as well as the inter- and the intra-species double occupancies vanish. This indicate that all atoms are localized and the local interactions prevent any double occupation. In this MIT,
all atoms of both species equally participate in the transition. Like in the three-component Hubbard model \cite{Gorelik,Inaba}, the MIT is referred to as the collective MIT \cite{Bo2013}. The filling $n_c=3/4$ can be considered as a particle-hole symmetry of filling $n_c=1/4$ \cite{Bo2013}. In this case, instead of the particle double occupancies, the hole double occupancies vanish. Since the hole double occupancies
$D_{cc}^{h}=D_{cc}+1-2 n_c$ and $D_{cf}^h=D_{cf}+ 1-n_f-n_c$, they indeed vanish for strong interactions $U_{cc}$, as one can see in Fig. \ref{fig2}.
For other filings $n_f$ we still observe the collective MIT when
the total particle filling is commensurate, i.e. $2 n_c+n_f=1$ or $2 n_c+ n_f=2$.

We summarize the characteristic features of the MIT, observed in the three-component FKM:
\begin{enumerate}
\item Collective MIT: $Z_{c}=Z_f=0$, $D_{cc}=D_{cf}=0$ (or $D_{cc}^h=D_{cf}^h=0$ for the hole case ).
\item Species-selective MIT: $Z_{c}=0$, $Z_f=1$, $D_{cc}=0$, $D_{cf}=n_c n_f$.
\item Inverse MIT: $Z_{c}=Z_f=0$, $D_{cc} \neq 0$, $D_{cf}=0$.
\end{enumerate}

The filling conditions of these MIT were previously found numerically by the DMFT \cite{Bo2013}. Within the mean-field slave boson approximation, we can derive them analytically.  The constrain (\ref{constraints1}), and the band renormalized factors of both atom species (\ref{rf}) in the mean-field approximation read
\begin{eqnarray}
e^2+2 p_c^2 + p_f^2 + 2 d_{cf}^2 + d_{\uparrow\downarrow}^2 + t^2 =1 , \\
Z_c = \frac{(e p_c + p_c d_{\uparrow\downarrow} + p_f d_{cf} + d_{cf} t )^2}{n_c(1-n_c)} , \\
Z_f = \frac{(e p_f + 2 p_c d_{cf} + d_{\uparrow\downarrow} t )^2}{n_f(1-n_f)} .
\end{eqnarray}
The particle fillings of both species can be also calculated analytically in the mean-field approximation
\begin{eqnarray}
n_c =  p_c^2 + d_{\uparrow\downarrow}^2 + d_{cf}^2 + t^2 , \\
n_f = p_f^2 + 2 d_{cf}^2 + t^2 .
\end{eqnarray}
In the collective Mott insulator $Z_{c}=Z_f=0$ and $D_{cc}= D_{cf} = 0$. These conditions lead to $e=d_{\uparrow\downarrow}=d_{cf}=t=0$. In this case, we obtain
\begin{eqnarray}
2 p_c^2 + p_f^2  =1 , \\
n_c =  p_c^2 , \\
n_f = p_f^2.
\end{eqnarray}
These mean-field equations indeed lead to $2 n_c + n_f=1$. This is the filling condition for the collective Mott transition at the commensurate fillings.

In the species-selective MIT $Z_{c}=0$, $Z_f=1$ and $D_{cc}=0$, and $D_{cf} \neq 0$.
Condition $D_{cc}=0$ leads to $d_{\uparrow\downarrow}=t=0$. Then we obtain the mean field equations
\begin{eqnarray}
e^2+2 p_c^2 + p_f^2 + 2 d_{cf}^2  =1  , \\
e p_c + p_f d_{cf} =0 , \\
(e p_f + 2 p_c d_{cf})^2 = n_f(1-n_f) , \\
n_c =  p_c^2  + d_{cf}^2 , \\
n_f = p_f^2 + 2 d_{cf}^2.
\end{eqnarray}
Without difficulty one can show that these mean-field equations lead to $e=p_f=0$, and $n_c=1/2$.

In the inverse MIT $Z_c=Z_f=0$, $D_{cf} = 0$, and $D_{cc} \neq 0$. These conditions lead to $e=p_c=d_{cf}=t=0$. In this case we obtain
\begin{eqnarray}
p_f^2 + d_{\uparrow\downarrow}^2 =1 , \\
n_c =  d_{\uparrow\downarrow}^2  , \\
n_f = p_f^2  .
\end{eqnarray}
These mean-field equations indeed lead to $n_c+n_f=1$. This is the filling condition for the inverse MIT.

So far we have analytically established the filling conditions for different MIT, which may occur in the three-component FKM.
From these analyses of the slave boson mean-field equations, one can see that in all insulating states there are no empty or triply occupied
sites. Since the considered mixture has two different double occupancies, there are only three possibilities of their vanishing. These three possibilities lead to three different kinds of the Mott insulator.
The collective Mott insulator is similar to the one in the three-component Hubbard model,
and is quite well studied in literature.~\cite{Gorelik,Inaba,Bo2013}
The species-selective and the inverse MIT exhibit special features in the band renormalization factors and the double occupancies of the atom species, that are not captured by the previous DMFT study~\cite{Bo2013}. Therefore we will study them in details.

\subsection{Species-selective and inverse metal-insulator transitions}

Since both the species-selective and the inverse MIT may occur at half filling $n_c=n_f=1/2$,
in this subsection we study the half filling case in details. In contrast to the single-band Hubbard model or the spinless FKM, where the half filling is commensurate with the particle component number, the half filling here is incommensurate.
The considered three-component FKM has two well-known limiting cases.
When $U_{cf}=0$, the two-component and single-component
atoms are completely decoupled. The two-component atoms just form the single-band
Hubbard model \cite{Hubbard}. The correlations between the atom components drive the mixture from metal to insulator.
Within the Kotliar-Ruckenstein slave boson mean-field approach,
the MIT occurs at the critical value
$U^C_{cc} = 16 \int\limits_{0}^{W/2}\varepsilon{\rho _0}\left( \varepsilon  \right)d\varepsilon = 2W$.~\cite{Kotliar} In the insulating state, both the band renormalization factor $Z_c$ and the
intra-species double occupancy $D_{cc}$ vanish.
When $U_{cc}=0$, the three-component FKM is equivalent to the spinless FKM \cite{FKM}. The inter-species local interaction drives the mixture from metal to insulator by splitting the two-component atom band.
A simple treatment within the Kotliar-Ruckenstein slave boson mean-field approach gives a continuous MIT at $U_{cf}^{C}=W$.
In the insulating state the inter-species double occupancy $D_{cf}$ vanishes while the intra-species double occupancy $D_{cc}$ remains finite.
When both local interactions $U_{cc}$ and $U_{cf}$ are finite, these MIT may occur, as we have discussed in the previous subsection.

\begin{figure}[b]
\begin{center}
\includegraphics[width=0.4\textwidth]{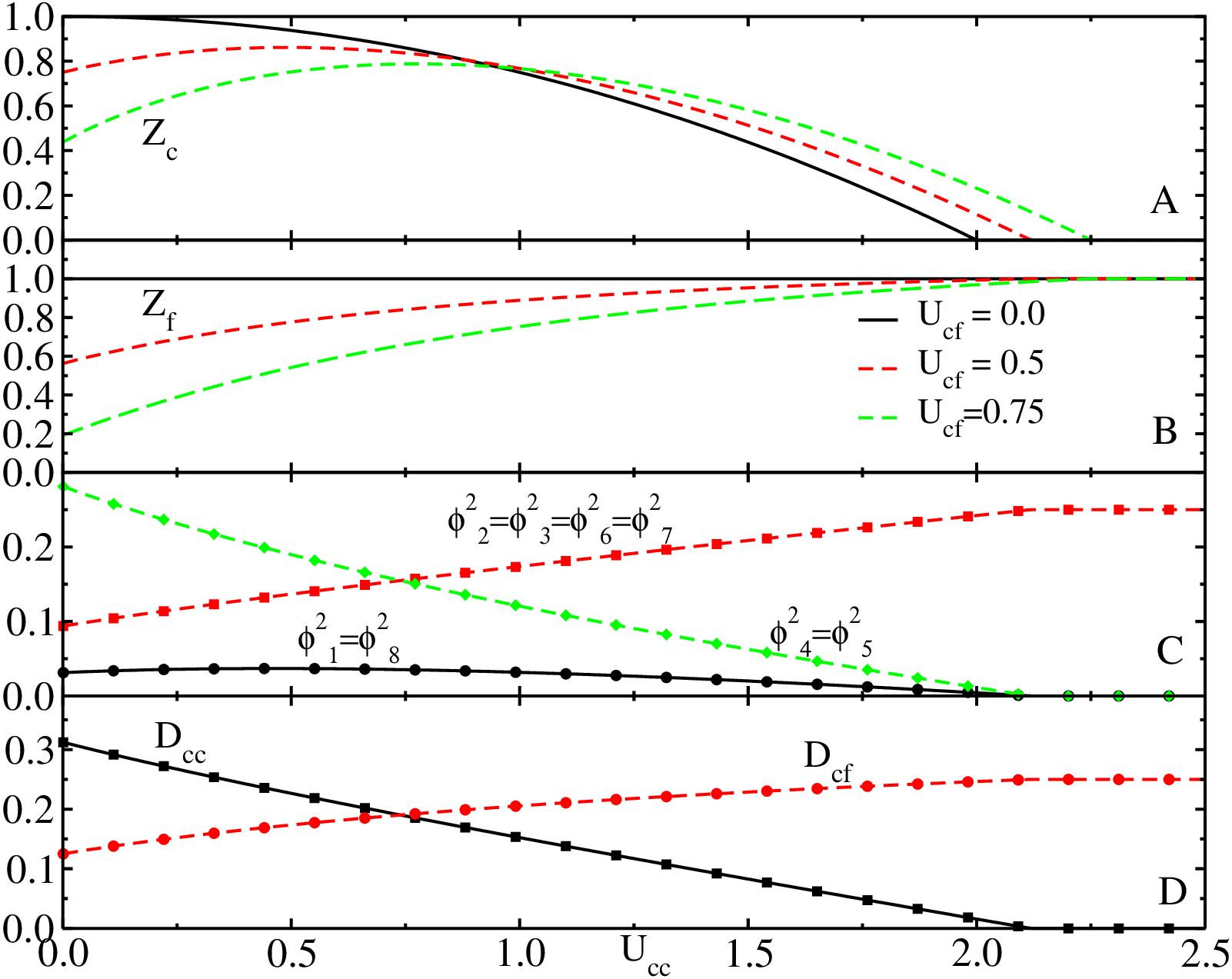}
\caption{(color online) The band renormalization factors $Z_c$ (panel A) and $Z_f$ (panel B), the boson condensation $\phi^2_{\Gamma}$ (panel C), and
the double occupancies $D_{cc}$ and $D_{cf}$ (panel D) as a function of the intra-species interaction $U_{cc}$ at half filling in the weak inter-species interaction region
($U_{cf}<W$).}
\label{Z_Ucc__halffilling_Ucf_small}
\end{center}
\end{figure}

\begin{figure}[t]
\begin{center}
\includegraphics[width=0.4\textwidth]{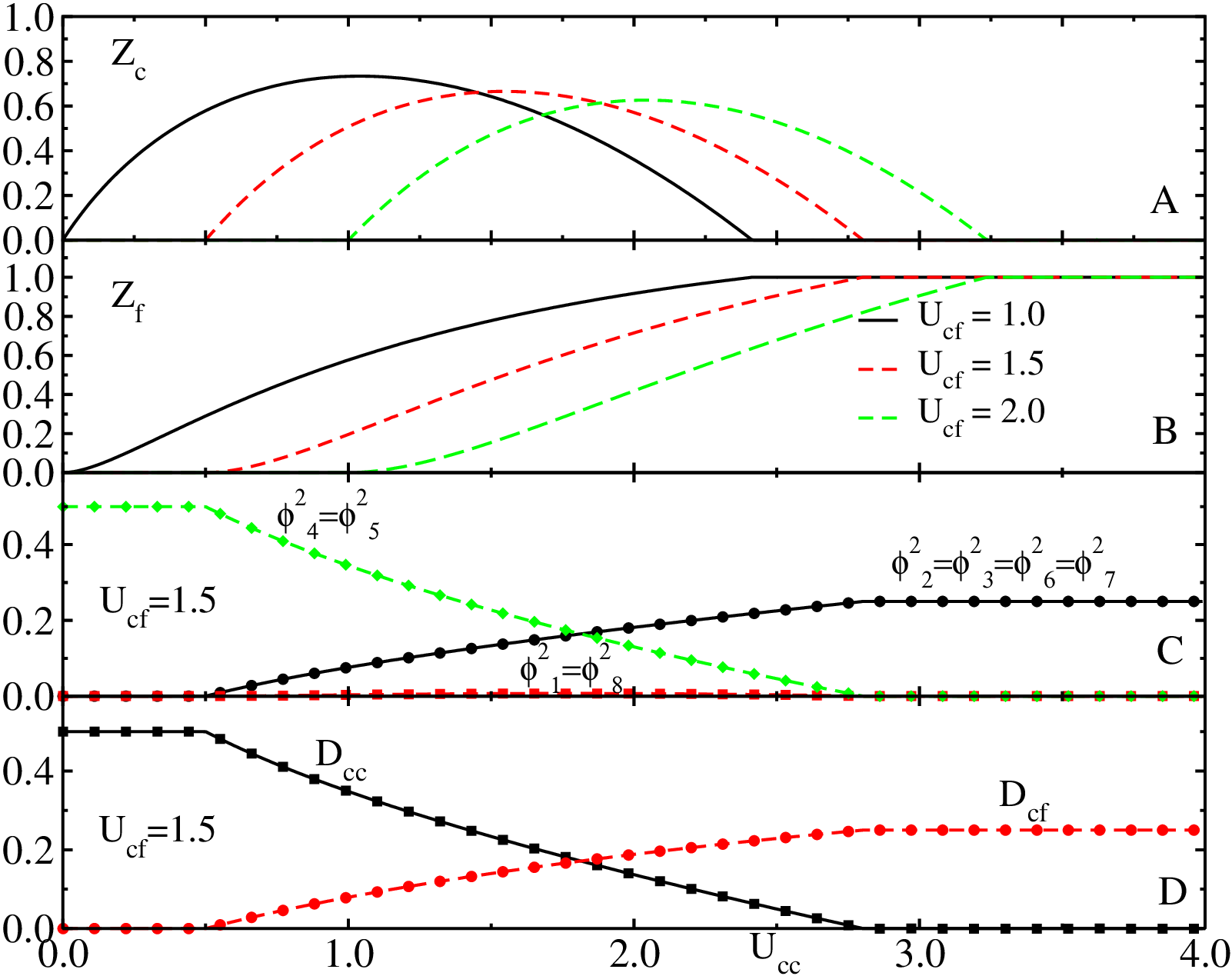}
\caption{(Color online) The band renormalization factors $Z_c$ (panel A) and $Z_f$ (panel B), the boson condensation $\phi^2_{\Gamma}$ (panel C), and
the double occupancies $D_{cc}$ and $D_{cf}$ (panel D) as a function of the intra-species interaction $U_{cc}$ at half filling in the strong inter-species interaction region ($U_{cf}>W$).}
\label{Z_Ucc__halffilling_Ucf_large}
\end{center}
\end{figure}
In Figs. \ref{Z_Ucc__halffilling_Ucf_small} and \ref{Z_Ucc__halffilling_Ucf_large}
we plot the band renormalization factors $Z_c$ and $Z_f$, and the double occupancies $D_{cc}$ and $D_{cf}$, and the boson condensation $\phi^2_{\Gamma}$ as a function of $U_{cc}$ for different values of $U_{cf}$.
In the region of weak inter-species interactions ($U_{cf}<W$), we observe only the species-selective MIT. In this MIT the band renormalization factor $Z_c$ and the double occupancy $D_{cc}$ of the two-component atoms vanish. However, the single-component atoms behave like the free ones in the insulating state. This MIT is similar to the one of the single band Hubbard model of the two-component atoms, despite the presence of the single-component atoms and their inter-species local interaction. For weak intra-species interaction $U_{cc}$,
the system is metallic, since the weak correlations cannot drive the system out of the metallic state. The local interactions only renormalize the effective mass of the two-component atoms and the energy level of the single-component atoms through the boson condensations.
Due to particle-hole symmetry, we observe that the following boson condensations are equal:
$e=t$, $p_c=d_{cf}$, and $d_{\uparrow\downarrow}=p_f$.
With further increasing $U_{cc}$, the boson condensations $e$, $t$,  $d_{\uparrow\downarrow}$, and $p_f$ decrease to zero value, as one can see in Fig.~\ref{Z_Ucc__halffilling_Ucf_small}. The vanishing of these boson condensations indicates the vanishing of the band renormalization factor $Z_c$ and the double occupancy $D_{cc}$ of the two-component atoms. It also leads to the localization of the two-component atoms. However, at the same time, the vanishing of these boson condensations also leads the band renormalization factor $Z_f$ of the single component to be free of interactions. It is a special feature of the species-selective MIT, and makes a distinction from the orbital-selective MIT \cite{Anisimov,Koga,Liebsch2004,Medici}. Since the boson condensations must be fixed by the constrain (\ref{constraints1}), the vanishing of the boson condensations $e$, $t$, $d_{\uparrow\downarrow}$, and $p_f$ results in a finite constant value of the boson condensations $p_c$ and $d_{cf}$. This leads the inter-species double occupancy $D_{cf}$ to reach the saturated value in the insulating state. The saturated value of the inter-species double occupancy is indeed the free one, since the single-component atoms are free of the interaction renormalization in the insulating state.

When $U_{cf}>W$, in contrast to the region $U_{cf}<W$, we observe two different MITs which occur with increasing $U_{cc}$. Figure \ref{Z_Ucc__halffilling_Ucf_large} shows the vanishing of the band renormalization factor $Z_c$ of the two-component atoms in weak and strong intra-species local interaction $U_{cc}$ regions. The boson condensations of the empty and triply-occupied states always vanish, except for the metallic region, where the band renormalization factor $Z_c$ is finite.
When the intra-species local interaction $U_{cc}$ is weak, the band renormalization factor $Z_f$ of the single-component atoms and the inter-species double occupancy $D_{cf}$ vanish too. It is the inverse MIT, where the intra-species local interaction drives the mixture from insulator to metal. In this insulating state, the boson condensations $p_c$ and $d_{cf}$ vanish, while the other boson condensations $p_f$ and $d_{\uparrow\downarrow}$ are constant. This leads the inter-species double occupancy to vanish, while the intra-species double occupancy maintains its constant value. With further increasing $U_{cc}$ the mixture is driven to the metallic state by correlations. In the metallic state, all boson condensations acquire finite values, that give finite values for the band renormalization factors $Z_c$ and $Z_f$ as well as the double occupancies $D_{cc}$ and $D_{cf}$. When the intra-species local interaction $U_{cc}$ is strong, only the band renormalization factor $Z_{cc}$ and the double occupancy $D_{cc}$ of the two-component atoms vanish. The band renormalization factor $Z_f$ of the single-component atoms and the inter-species double occupancy $D_{cf}$  reach their free saturated values. This is the species-selective MIT, which we have discussed previously.

\begin{figure}[b]
\begin{center}
\includegraphics[width=0.4\textwidth]{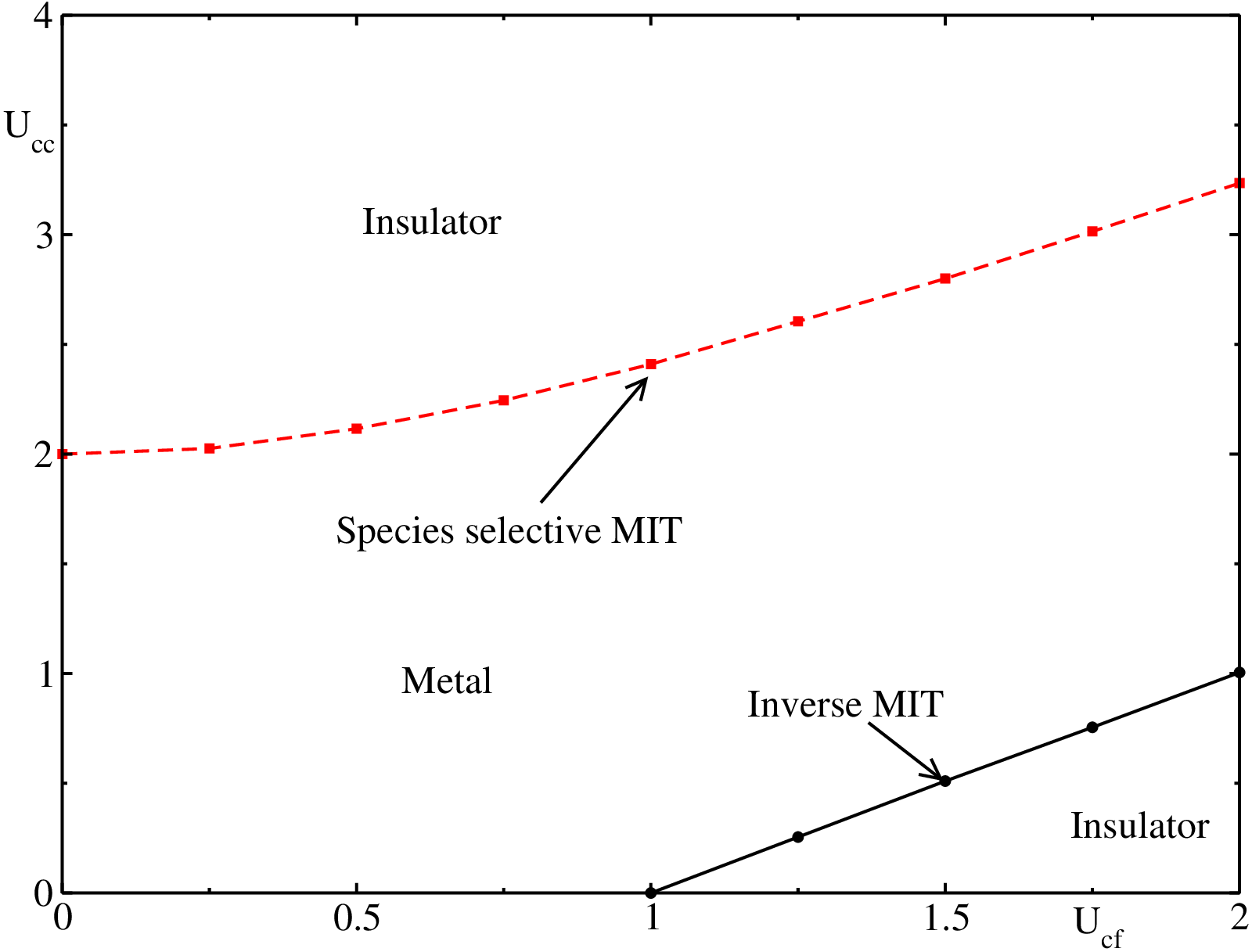}
\caption{(Color online) Phase diagram at half-filling $n_{c}=n_f=1/2$.}
\label{phase_half_filling}
\end{center}
\end{figure}

The above results are summarized  into the phase diagram plotted in Fig.
\ref{phase_half_filling}.
It shows possibilities of two distinct MITs at half filling. One MIT is species selective, and the other is the inverse one. The phase diagram agrees well with the one obtained by the DMFT \cite{Bo2013},
except for the first-order phase transition of the species-selective MIT.
The Kotliar-Ruckenstein slave boson mean-field approximation usually overestimates the critical interaction values of the Mott transitions.
For instance, for the single-band Hubbard model, the critical interaction value of the Mott transition calculated by the Kotliar-Ruckenstein slave boson mean-field approximation is almost twice as larg as that obtained by the DMFT \cite{GKKR,Kotliar}.
The two obtained MITs occur in different regions of the parameters. The species-selective MIT occurs in the region of
strong intra-species local interactions ($U_{cc} \geq 2 W$) and the inverse MIT occurs in the region of weak ones ($U_{cc} < 2W$).
The critical value $U_{cc}^{\rm ss}$ of the species-selective MIT monotonously increases with $U_{cf}$. Therefore at a given strong $U_{cc}$, the inter-species local interaction drives the mixture from insulator to metal.
The critical value $U^{\rm inv}_{cc}$ of the inverse MIT is indeed linear in $U_{cf}$. It vanishes at
$U_{cf}=W$. For weak inter-species local interactions, i.e., $U_{cf}<W$, we cannot observe the inverse MIT.

Within the Kotliar-Ruckenstein slave boson mean-field approximation, the critical value of $U_{cc}$ can also analytically be obtained. For the species-selective MIT, as we have analyzed in the previous section, when the transition approaches $e \rightarrow 0$, $p_f \rightarrow 0$ at $d_{\uparrow\downarrow}=t=0$. Then nearby the transition point, the ground-state energy can be expanded up to second order of $e$ and $p_f$
\begin{eqnarray}
E &=& - W (e+p_f)^2 + U_{cc}(p_f^2+e^2) \nonumber \\
&& + U_{cf} (\frac{1}{2} + e^2 - p_f^2) .
\end{eqnarray}
Minimizing this ground-state energy with  respect to $e$ and $p_f$, and then taking the limit $e \rightarrow 0$, $p_f \rightarrow 0$, we obtain
the critical value of the species-selective MIT
\begin{eqnarray}
U_{cc}^{\rm ss} = W + \sqrt{W^2 + U_{cf}^2} .
\end{eqnarray}
For the inverse MIT, when the transition approaches $e \rightarrow 0$, $p_c \rightarrow 0$ at $d_{cf}=t=0$.  In a similar way, we obtain the critical value of the inverse MIT
\begin{eqnarray}
U_{cc}^{\rm inv} = W + U_{cf} .
\end{eqnarray}
One can see that $U_{cc}^{\rm ss}$ is always larger than $U_{cc}^{\rm inv}$. This means that the inverse MIT always occurs before the species-selective one. Therefore, for strong inter-species interaction ($U_{cf}>W$), there would be a reentrant effect of MIT. With increasing the intra-species interaction $U_{cc}$, the mixture first stays in the insulating state, then it is transformed into the metallic state when
$U_{cc} > U_{cc}^{\rm inv}$, and finally, the mixture goes back to the insulating state when $U_{cc} > U_{cc}^{\rm ss}$. Note that the first and second insulating states are quite different. In the first insulating state, the inter-species double occupancy vanishes, while the intra-species one remains finite. In the second insulating state, the single-component atoms become free, albeit being localized, and the intra-species double occupancy vanishes.
Experiments would observe this reentrant effect of MITs by measuring the intra- and inter-species double occupancies.

\section{\label{sec4} Conclusion}

We have studied the MIT in the three-component FKM by the Kotliar-Ruckenstein slave boson approach at the mean-field level. Although the slave boson mean field approximation is simple, the obtained results reproduce all important features of the MIT, that are obtained within the more sophisticated DMFT. In particular, the filling conditions and the critical value of the local interaction of the MIT are analytically established. Moreover, the slave boson mean-field approximation allows us to clarify the nature of different MITs, which occur in the three-component FKM, in a simple physics picture. This is an advantage of the slave boson approach. In the collective Mott insulator, all band renormalization factors as well as the double occupancies vanish. The collective MIT occurs only at the commensurate fillings. At the incommensurate fillings, the species-selective or the inverse MIT may occur. The species-selective MIT occurs only at half filling of the two-component atoms irrespective of the filling of the single-component atoms. Actually, in the species-selective MIT only two-component atoms are involved, while the single-component ones become free of interactions. The freedom of the single-component atoms in the species-selective Mott insulator is a special feature that makes its distinction from other Mott insulators.
The inverse MIT occurs when the total filling of the single-component atoms and of one component of the two-component atoms reaches a unit. This MIT occurs only for weak intra-species local interactions. The obtained results, which are in good agreement with the DMFT, suggest that the Kotliar-Ruckenstein slave boson at the mean-field level is already adequate to describe the MIT in multicomponent fermion-fermion mixtures of ultracold atoms. In addition, the slave boson mean-field approximation also provides the criteria of the MIT through the band renormalization factors and the double occupancies.
Since experiments of ultracold atom mixtures detect the Mott insulator by counting the number of doubly occupied sites, it is a challenge to observe the MIT in multicomponent fermion-fermion mixtures of ultracold atoms.

\begin{acknowledgments}
We would like to thank Yoshiro Takahashi for helpful
discussions on experimental realizations of the three-component
Falicov-Kimball model in optical lattices. This work was financially supported by the National Foundation for Science and Technology Development
 of Vietnam under Grant No. 103.01-2014.23.
\end{acknowledgments}

\end{document}